\documentclass[fleqn,10pt]{wlscirep}
\usepackage[utf8]{inputenc}
\usepackage[T1]{fontenc}
\usepackage{graphicx,epsf,bm,amsmath,amsfonts,amssymb,epstopdf,hyperref,color,verbatim,multirow,bm}
\hypersetup{colorlinks=true,urlcolor=blue,citecolor=blue,linkcolor=blue,menucolor=blue,anchorcolor=blue,filecolor=blue}

\title{Implications of cosmologically coupled black holes for pulsar timing arrays}

\author[1,2]{Marco Calz\`{a}}
\author[1]{Francesco Gianesello}
\author[1,2]{Massimiliano Rinaldi}
\author[1,2*]{Sunny Vagnozzi}
\affil[1]{Department of Physics, University of Trento, Via Sommarive 14, 38123 Povo (TN), Italy}
\affil[2]{Trento Institute for Fundamental Physics and Applications-INFN, Via Sommarive 14, 38123 Povo (TN), Italy}

\affil[*]{sunny.vagnozzi@unitn.it}


\begin{abstract}
It has been argued that realistic models of (singularity-free) black holes (BHs) embedded within an expanding Universe are coupled to the large-scale cosmological dynamics, with striking consequences, including pure cosmological growth of BH masses. In this pilot study, we examine the consequences of this growth for the stochastic gravitational wave background (SGWB) produced by inspiraling supermassive cosmologically coupled BHs. We show that the predicted SGWB amplitude is enhanced relative to the standard uncoupled case, while maintaining the $\Omega_{\text{gw}} \propto f^{2/3}$ frequency scaling of the spectral energy density. For the case where BH masses grow with scale factor as $M_{\text{bh}} \propto a^3$, thus contributing as a dark energy component to the cosmological dynamics, $\Omega_{\text{gw}}$ can be enhanced by more than an order of magnitude. This has important consequences for the SGWB signal detected by pulsar timing arrays, whose measured amplitude is slightly larger than most theoretical predictions for the spectrum from inspiraling binary BHs, a discrepancy which can be alleviated by the cosmological mass growth mechanism.
\end{abstract}

\begin{document}

\flushbottom
\maketitle

\thispagestyle{empty}

\section*{Introduction}
\label{sec:introduction}

One of the most significant open questions in physics is to determine the origin of the late-time accelerated expansion of the Universe, first inferred in 1998, and now directly or indirectly confirmed by a wide variety of probes~\cite{SupernovaSearchTeam:1998fmf,SupernovaCosmologyProject:1998vns}. A number of models ascribe this to either a dark energy (DE) component, which could be the manifestation of one or more new particles, or to the effects of large-scale modifications of General Relativity~\cite{Sotiriou:2008rp,Clifton:2011jh,Sebastiani:2016ras,Nojiri:2017ncd,Huterer:2017buf}. Nevertheless, a compelling model of DE remains elusive, and this has triggered various novel lines of investigation, where cosmic acceleration arises as a result of more exotic effects. In an intriguing line of research, a potential solution to the problem of cosmic acceleration is connected to another major problem in theoretical physics: the inevitable but undesirable existence of singularities following continuous gravitational collapse~\cite{Penrose:1964wq,Hawking:1970zqf,Sebastiani:2022wbz}. Could the physics of black holes (BHs), and in particular the mechanism eventually responsible for the resolution of BH singularities, be connected to the physics of cosmic acceleration?

Attempts to embed BHs within an expanding Universe, thereby subject to cosmological boundary conditions with a time-dependent geometry, have been ongoing since the early 1930s~\cite{McVittie:1933zz,Nolan:1993ghw,Faraoni:2007es,Kaloper:2010ec,Lake:2011ni,Faraoni:2012gz,daSilva:2012nh,Croker:2019kje,Croker:2020plg,Croker:2022vdq}. The idea is that there should be a global solution to the Einstein equations that can simultaneously describe a BH at small scales and an expanding Universe at large scales. Surprising features characterize the resulting objects, which appear to be singularity-free and are generally coupled to the large-scale cosmological dynamics: we shall hence refer to them as \textit{cosmologically coupled BHs}. In particular, there are known cosmologically coupled BHs which expand hand-in-hand with the embedding cosmology, and whose mass increases with the expanding Universe independently of accretion and mergers~\cite{Faraoni:2007es,Cadoni:2023lum}. In fact, the common assumption of static black hole horizons existing in time-dependent metrics bears unphysical consequences, such as naked singularities and geodesic incompleteness~\cite{Faraoni:2024ghi}, leading to the conclusion that BH event horizons have to be cosmologically coupled. Interest in the topic of cosmologically coupled BHs has been largely theoretical so far, since one can argue that separation of scales justifies considering BHs as asymptotically flat and decoupled from the cosmological expansion. However, when dealing with supermassive BHs which have existed for a long time span, cosmological coupling effects can become relevant. 

Intriguingly, possible signs of pure cosmological mass growth were recently searched for and reported in the growth of supermassive BHs hosted by red-sequence elliptical galaxies (where growth by accretion and mergers is expected to be insignificant), with the preferred rate of mass growth in terms of the cosmological scale factor being consistent with $M_{\text{bh}} \propto a^3$~\cite{Farrah:2023opk}. Since their resulting physical density is constant, these cosmologically coupled BHs would then contribute to the cosmological dynamics as a DE component, and this intriguing possibility has spurred various follow-up theoretical and observational studies~\cite{Rodriguez:2023gaa,Parnovsky:2023wkc,Avelino:2023rac,Andrae:2023wge,Lei:2023mke,Sadeghi:2023cpd,Ghodla:2023iaz,Garcia-Bellido:2023exi,Amendola:2023ays,Gaur:2023hmk,Cadoni:2023lqe,Deliduman:2023caa,Lacy:2023kbb,Christiansen:2024kxn,Croker:2024jfg,Cadoni:2024jxy}. However, a putative cosmological coupling of BH masses would also have other important observational consequences. In particular, cosmologically coupled BHs are expected to inspiral and undergo (binary) mergers throughout their history, resulting in the emission of gravitational waves (GWs), whose incoherent superposition would result in a stochastic GW background (SGWB) signal. These considerations become all the more relevant in light of the SGWB signal for which evidence was recently provided by various pulsar timing arrays (PTAs), including NANOGrav~\cite{NANOGrav:2023gor}, EPTA~\cite{EPTA:2023fyk}, PPTA~\cite{Reardon:2023gzh}, and CPTA~\cite{Xu:2023wog}. The origin of this signal (and of the earlier NANOGrav 12.5-year signal) remains unknown, although several proposals, both astrophysical and cosmological (primordial) in origin, have been put forward~\cite{Ellis:2020ena,Blasi:2020mfx,DeLuca:2020agl,Vagnozzi:2020gtf,Li:2020cjj,Benetti:2021uea,Vagnozzi:2023lwo,Oikonomou:2023qfz,Huang:2023chx,Wang:2023ost,Ghosh:2023aum,Figueroa:2023zhu,Wang:2023sij,Choudhury:2023kam,Jiang:2023gfe,Choudhury:2023hfm,Oikonomou:2023bli,Chowdhury:2023xvy}. A natural question to ask is therefore what is the SGWB signal resulting from a population of inspiraling, cosmologically coupled, supermassive BHs.

The aim of this pilot study is to take a first step towards working out the aforementioned signal. This allows us to establish yet another observational channel towards testing cosmological couplings between BHs and the large-scale cosmological expansion, and ultimately a potential mechanism for singularity resolution. Indeed, the cosmological coupling allows to bypass the Penrose-Hawking singularity theorems and can lead to BH solutions that are non-singular, even in the context of standard General Relativity~\cite{Cadoni:2022chn,Cadoni:2023lum}. To work out the SGWB signal from inspiraling cosmologically coupled BHs, we follow the seminal work of Phinney~\cite{Phinney:2001di}, revisited in light of a phenomenological parametrization for the cosmological mass growth mechanism resulting from the cosmological coupling~\cite{Croker:2019mup,Farrah:2023opk}. Intriguingly, while astrophysical models of inspiraling supermassive BH binaries are broadly able to reproduce the signal observed by PTAs, its amplitude is slightly high compared to the expectations from virtually all models. We show that, under a very minimal set of hypotheses, inspiraling cosmologically coupled BHs can naturally boost the amplitude of the SGWB signal, while keeping the spectral index unaffected, thus maintaining the same $\Omega_{\text{gw}} \propto f^{2/3}$ frequency scaling of the SGWB spectral energy density parameter.

The rest of this work is then organized as follows. We briefly review the standard computation of the SGWB resulting from inspiraling uncoupled BH binaries. The calculation is then generalized to allow for cosmologically coupled BHs. Afterwards we provide a critical discussion of our results, focusing in particular on the amplitude of the SGWB signal and contrasting it against the NANOGrav signal, before drawing concluding remarks. Throughout our work, we set $c=1$, but for clarity we keep $G$ in our formulae.

\section*{Stochastic gravitational wave background from inspiraling black hole binaries}
\label{sec:binaries}

Of interest to the present work is the stochastic gravitational wave background generated by a population of BH binaries inspiraling due to gravitational radiation losses, assuming time-dependent masses provided purely by  cosmological mass growth as expected within the classes of solutions discussed in the Introduction. To do so, we modify the work of Phinney~\cite{Phinney:2001di}, by introducing $z$-dependent masses for the components of the binaries. The main quantity of interest to PTA experiments, such as NANOGrav, is the timing residual power spectral density $\Phi(f)$, itself related to the SGWB strain power spectrum $h_c(f)$, where $f$ denotes frequency. The SGWB strain power spectrum is related to the SGWB spectral energy density parameter $\Omega_{\text{gw}}(f)$, which measures the present-day SGWB energy density per logarithmic frequency range, in units of the critical density of the Universe $\rho_c$. Denoting by $\rho_{\text{gw}}(f)$ the present-day GW energy density, $\Omega_{\text{gw}}(f)$ is given by the following:
\begin{eqnarray}
\Omega_{\text{gw}}(f)= \frac{1}{\rho_c}\frac{d \rho_{GW}}{d \log(f)}\,,
\label{eq:omegagw}
\end{eqnarray}
where the critical density is given by $\rho_c=3H_0^2/8\pi G$, with $H_0$ being the Hubble constant. Following the same steps as the Phinney's work~\cite{Phinney:2001di} (steps which, so far, are unmodified by the cosmological mass growth mechanism, as long as this does not alter the amount by which the energy/frequency of gravitons redshifts with the expansion of the Universe), the SGWB spectral energy density parameter can then be expressed as:
\begin{eqnarray}
\Omega_{\text{gw}}(f)=\frac{1}{\rho_c}\int_{0}^{\infty} dz\,\frac{n(z)}{1+z} \frac{dE_{\text{gw}}}{d\log f_r} \Bigg\vert_{f_r=f(1+z)}\,,
\label{eq:omegagwphinneystandard}
\end{eqnarray}
where $n(z)$ indicates the comoving number density of events (number of mergers/remnants per unit comoving volume), whereas $f$ and $f_r$ are the frequency of the SGWB measured on Earth and in the source's cosmic rest frame respectively and are related by $f_r=f(1+z)$. Finally $E_{\text{gw}}(f)$ measures the total energy emitted in GWs in a certain frequency band, and its functional form depends on the specific population of sources/events producing the SGWB.

We now consider a binary system of supermassive BHs in circular orbit and with component masses $M_1$ and $M_2$. We assume that the BHs inspiral, and eventually merge, uniquely driven by gravitational radiation loss. For such a system, we have~\cite{Thorne:1987ghw,Phinney:2001di}:
\begin{eqnarray}
\frac{dE_{\text{gw}}}{d\log f_r}=f_r\frac{dE_{\text{gw}}}{df_r}=\frac{\pi^{2/3}G^{2/3}\mathcal{M}^{5/3}}{3}f_r^{2/3}\,,
\label{eq:dedfr}
\end{eqnarray}
where the so-called chirp mass ${\cal M}$ is given by the following:
\begin{eqnarray}
\mathcal{M}= \left [ \frac{M_1 M_2}{(M_1 + M_2)^{1/3}} \right ] ^{\frac{3}{5}}\,.
\label{eq:chirpmass}
\end{eqnarray}
Combining the above equations one finds, for the standard case with no cosmological mass growth~\cite{Phinney:2001di}:
\begin{eqnarray}
\Omega_{\text{gw}}(f)= \frac{8(\pi G \mathcal{M})^{5/3}}{9H_0^2}f^{2/3}\int_{z_{\min}}^{z_{\max}}dz\,\frac{n(z)}{(1+z)^{1/3}}\,.
\label{eq:omegagwstandard}
\end{eqnarray}
It proves convenient to introduce the quantity $n_0=\int_0^\infty dz\,n(z)$, which quantifies the present-day comoving number density of merged remnants, as well as the merger rate per comoving volume $\dot{n}(z)$, such that $n(z)=\dot{n}(z)dt/dz$. We further define the following quantity:
\begin{eqnarray}
\langle (1+z)^{-1/3} \rangle\equiv\frac{1}{n_0}\int_{z_{\min}}^{z_{\max}}dz\,\frac{n(z)}{(1+z)^{1/3}}=\frac{1}{H_0n_0}\int_{z_{\min}}^{z_{\max}}dz\,\frac{\dot{n}(z)}{(1+z)^{4/3}E(z)}\,,
\label{eq:averageredshift}
\end{eqnarray}
where in the second equality we have made use of the time-redshift relation in a Friedmann-Lema\^{i}tre-Robertson-Walker Universe, and $E(z) \equiv H(z)/H_0$ is the unnormalized expansion rate, which in a $\Lambda$CDM Universe is given by the following:
\begin{eqnarray}
E(z)= \sqrt{\Omega_m(1+z)^3+\Omega_r(1+z)^4+\Omega_{\Lambda}}\,,
\label{eq:ez}
\end{eqnarray}
with $\Omega_m$, $\Omega_r$, and $\Omega_{\Lambda}$ denoting the matter, radiation, and cosmological constant density parameters respectively (although we note that the radiation component is negligible at the late times of interest to us). With these definitions, $\Omega_{\text{gw}}(f)$ can then be written as follows:
\begin{eqnarray}
\Omega_{\text{gw}}=\frac{8(\pi G\mathcal{M})^{5/3}n_0}{9H_0^2}f^{2/3} \langle (1+z)^{-1/3} \rangle \,,
\label{eq:omegagwstandardaverageredshift}
\end{eqnarray}
within the range of validity $f_{\min}<f<f_{\max}$ (relative to the source rest frame), where the lower and upper limits are set by the separation of the system at birth and the final orbit prior to Roche contact respectively (the latter is a proxy for the end of the inspiraling phase), with both limits mapped into the minimum and maximum redshifts $z_{\min}$ and $z_{\max}$ through $f_r=f(1+z)$. In Eq.~(\ref{eq:omegagwstandardaverageredshift}) we observe the well-known $f^{2/3}$ scaling of the spectral energy density parameter for the SGWB produced by inspiraling binary BHs.

The results of GW searches, including SGWB searches from PTA experiments, are often reported in terms of the power spectrum of the characteristic GW strain $h_c$, itself related to the SGWB spectral energy density parameter as follows:
\begin{eqnarray}
h_c^2(f)=\frac{4G\rho_c}{\pi}f^{-2}\Omega_{\text{gw}}(f)= \frac{3H_0^2}{2\pi^2}f^{-2}\Omega_{\text{gw}}(f)\,,
\label{eq:hc2omegagw}
\end{eqnarray}
which, combined with Eq.~(\ref{eq:omegagwstandardaverageredshift}), leads to the following expression:
\begin{eqnarray}
h_c^2(f) = \frac{4(G\mathcal{M})^{5/3}n_0}{3\pi^{1/3}}f^{-4/3} \langle (1+z)^{-1/3}\rangle \,,
\label{eq:hc2f}
\end{eqnarray}
Of particular interest to PTA experiments is the power spectral density of the timing residuals $\Phi(f)$, related to the power spectrum of the characteristic SGWB strain (and therefore to the SGWB spectral energy density parameter) by the following:
\begin{eqnarray}
\Phi(f)=\frac{1}{12\pi^2}f^{-3}h_c^2(f)=\frac{H_0^2}{8\pi^4}f^{-5}\Omega_{\text{gw}}(f)\,.
\label{eq:phif}
\end{eqnarray}
It is customary to normalize $\Phi(f)$ relative to a reference frequency $f_{\text{ref}}$, usually either $1\,{\text{yr}}^{-1}$ or $(10\,{\text{yr}})^{-1}$, as follows:
\begin{eqnarray}
\Phi(f)=\frac{A_{\text{ref}}^2}{12\pi^2} \left ( \frac{f}{f_{\text{ref}}} \right ) ^{-\gamma} f_{\text{ref}}^{-3}\,.
\label{eq:phiffref}
\end{eqnarray}
Combining Eqs.~(\ref{eq:omegagwstandardaverageredshift},\ref{eq:hc2f},\ref{eq:phif}) we see that the idealized case (in the absence of environmental effects -- we will return to this point later) corresponds to the well-known value for the spectral index $\gamma=5-2/3=13/3$, which is indeed very often taken as a benchmark value in PTA SGWB searches. Finally, once more combining Eqs.~(\ref{eq:omegagwstandardaverageredshift},\ref{eq:hc2f},\ref{eq:phif}), we find that the amplitude $A_{\text{ref}}$ in Eq.~(\ref{eq:phiffref}) is given by the following expression:
\begin{eqnarray}
A_{\text{ref}}^2= \frac{4(G\mathcal{M})^{5/3}n_0}{3\pi^{1/3}} \langle (1+z)^{-1/3}\rangle = \frac{4(G \mathcal{M})^{5/3}}{3\pi^{1/3}H_0}\int_{z_{\min}}^{z_{\max}}dz\,\frac{\dot{n}(z)}{(1+z)^{4/3}E(z)} \,.
\label{eq:a}
\end{eqnarray}
In what follows, we will implicitly set the reference frequency to $f_{\text{ref}}=1\,{\text{yr}}^{-1} \approx 3.17 \times 10^{-8}\,{\text{Hz}}$, and to lighten the notation we simply denote $A_{\text{ref}}=A_{\text{yr}}=A$.

\section*{Inspiraling cosmologically coupled black holes}
\label{sec:cosmologicallycoupledbhs}

Up to now, we have not made any non-standard assumption concerning the BHs whose inspiraling produces the SGWB signal with amplitude characterized by Eq.~(\ref{eq:a}). We now posit that the BHs in question are non-singular, cosmologically coupled vacuum energy objects, whose mass is subject to cosmological growth. We recall once more that such a behaviour, while surprising, enjoys strong theoretical motivation and is somewhat inevitable if one wishes to embed BH solutions within cosmological boundary conditions with a time-dependent geometry~\cite{McVittie:1933zz,Nolan:1993ghw,Faraoni:2007es,Kaloper:2010ec,Lake:2011ni,Faraoni:2012gz,daSilva:2012nh,Croker:2019kje,Croker:2020plg,Croker:2022vdq}. Although explicit cosmologically coupled solutions are lacking, we follow a model-agnostic phenomenological approach adopted earlier~\cite{Croker:2021duf}, including in the work providing possible evidence for purely cosmological mass growth from observations of supermassive BHs hosted by red-sequence elliptical galaxies~\cite{Farrah:2023opk}. We parametrize the evolution of a BH mass as a function of redshift $z$ as follows:
\begin{eqnarray}
M(z)=M_i\Theta(z-z_i)+M_i \left ( \frac{z_i +1}{z+1} \right ) ^k \Theta(z_i-z)\,,
\label{eq:mz}
\end{eqnarray}
where $\Theta$ is the usual Heaviside step function. Effectively, Eq.~(\ref{eq:mz}) describes a BH whose mass $M$ is constant up to a certain transition redshift $z_i$ when the BH becomes cosmologically coupled, at which point its mass scales with the scale factor $a$ as a power-law, $M \propto a^k$, where $k$ is the coupling strength. We note that, because cosmological mass and number densities for non-relativistic objects scale as $a^{-3}$, the case $k=3$ (which is also the maximum allowed value of $k$ consistent with causal material with positive energy density) where BHs gain mass as $M \propto a^3$ leads to their physical densities being constant (in the absence of production), and therefore to their entering the Friedmann equations as an effective vacuum energy component. It has indeed been shown that vacuum energy interior solutions with cosmological boundaries lead to $k \sim 3$~\cite{Croker:2019mup}. We remark that, while phenomenological, our parametrization in Eq.~(\ref{eq:mz}) nevertheless enjoys theoretical motivation.

In a binary system whose masses $M_1$ and $M_2$ evolve as in Eq.~(\ref{eq:mz}), with $z_i$ being an universal parameter and therefore equal for both components, the same evolution clearly characterizes the chirp mass -- in other words, we can replace $M \to {\cal M}$ in Eq.~(\ref{eq:mz}). In the derivation of the SGWB spectral energy density parameter, the key equation which then gets modified is Eq.~(\ref{eq:dedfr}), since the energy emitted in GWs is altered as the masses of the two components grow. In particular, Eq.~(\ref{eq:dedfr}) is then modified to the following:
\begin{eqnarray}
\frac{dE_{\text{gw}}}{d\log f_r}=\frac{\pi^{2/3}G^{2/3}\mathcal{M}_i^{5/3}}{3}f_r^{2/3} \left [ \Theta(z-z_i) + \Theta(z_i-z) \left ( \frac{z_i+1}{z+1} \right ) ^{5k/3} \right ] \,.
\end{eqnarray}
Motivated by both phenomenological and observational considerations, we now make the reasonable assumption that the redshift at which the coupling switches on satisfies $z_{\min},z_{\max}<z_i$. Indeed, if cosmologically coupled BHs are to play a role in the DE phenomenon, then the cosmological coupling must have already been active, at the very least, at the onset of DE domination ($z \simeq 0.7$), but likely much earlier. Moreover, observations of supermassive BHs within elliptical galaxies are consistent with the cosmological coupling being in place at redshifts as high as $2.5$~\cite{Farrah:2023opk}, from which it follows that $z_i \gtrsim 2.5$. On the other hand, the sources of interest to PTAs are high mass binaries at low-to-moderate redshifts ($z \lesssim 2$). Given that our work is centered around the SGWB detected by PTAs, we can therefore safely assume $z_{\min},z_{\max}<z_i$. In other words, it is reasonable to assume that the coupling is active throughout the whole redshift range of interest to PTAs. Under this assumption, the SGWB spectral density parameter is then given by the following:
\begin{eqnarray}
\Omega_{\text{gw}} = \frac{8(\pi G\mathcal{M})^{5/3}}{9H_0^3}f^{2/3}(z_i+1)^{5k/3}\int_{z_{\min}}^{z_{\max}}dz\,\frac{\dot{n}(z)}{(1+z)^{(5k+4)/3}E(z)} \,.
\label{eq:omegagwcosmologicallycoupled}
\end{eqnarray}
It is worth noting that the modification we introduced has altered the amplitude of the SGWB signal, but not its frequency dependency, which remains $\Omega_{\text{gw}} \propto f^{2/3}$. Since the amplitude of $\Omega_{\text{gw}}$ is modified, we can expect the same to occur for the amplitude of the timing residuals power spectral density. In fact, retracing the steps which led to Eq.~(\ref{eq:a}), we find that the (squared) amplitude is modified to the following:
\begin{eqnarray}
A^2=\frac{4(G\mathcal{M})^{5/3}}{3\pi^{1/3}H_0} (z_i+1)^{5k/3}\int_{z_{\min}}^{z_{\max}}dz\,\frac{\dot{n}(z)}{(1+z)^{(5k+4)/3}E(z)} \,.
\label{eq:amodified}
\end{eqnarray}
We note that the cosmological mass growth does not alter the frequency scaling of $\Phi(f)$, which remains of the $\Phi(f) \propto f^{-13/3}$ form, just as in the standard (idealized) case.

\section*{Results and discussion}
\label{sec:results}

\begin{figure*}[!t]
\centering
\includegraphics[width=0.55\linewidth]{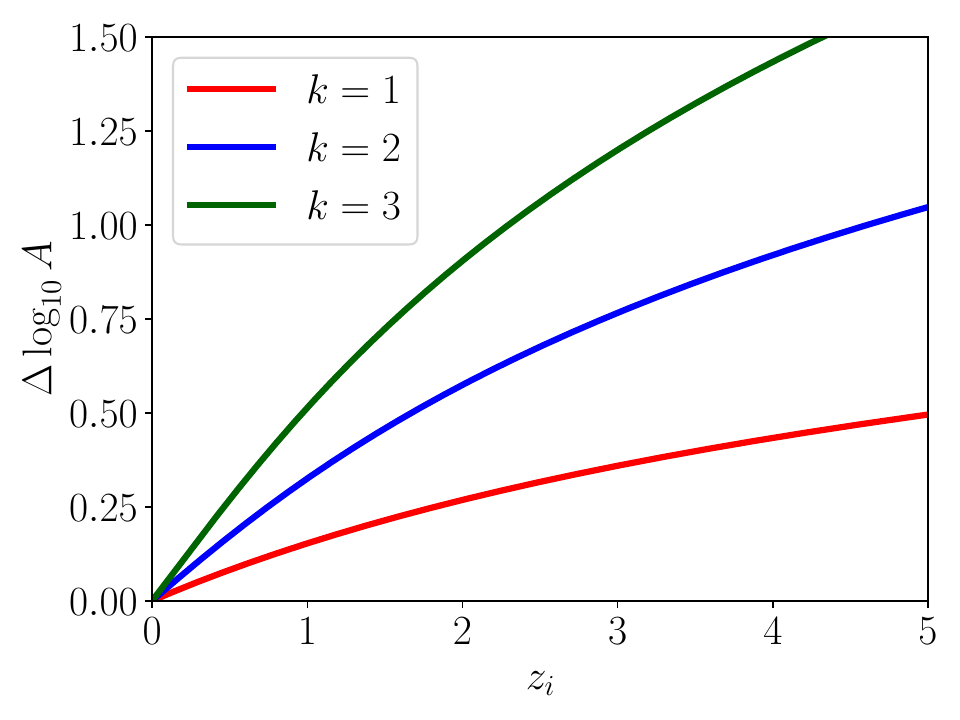}\,\,\,\,\,\includegraphics[width=0.41\linewidth]{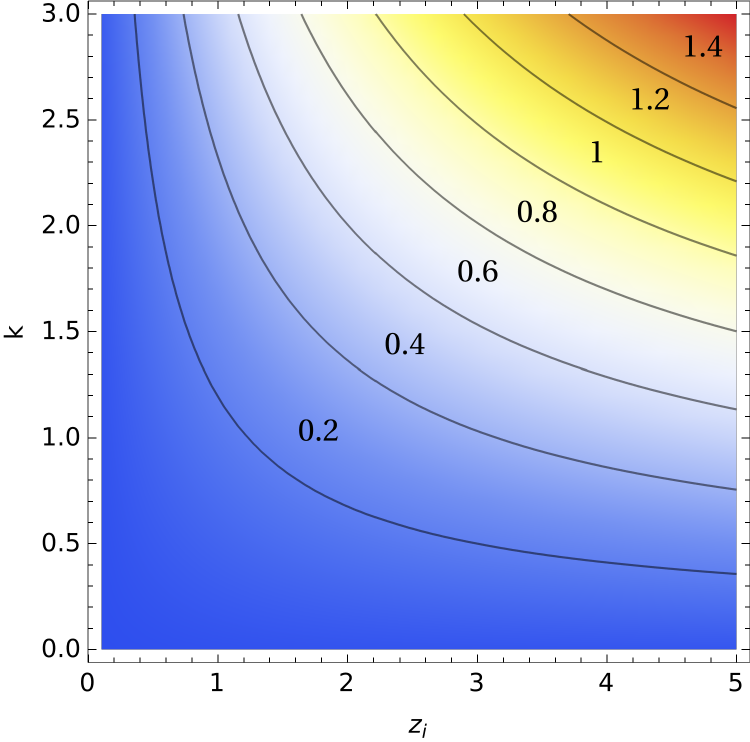}
\caption{\textit{Left panel}: logarithmic amplitude shift to the amplitude of the SGWB from inspiraling supermassive BHs in the presence of the cosmological coupling mechanism studied in this work, $\Delta\log_{10}A$, as a function of the transition redshift $z_i$ at which the coupling becomes effective, and for three representative values of the coupling strength $k$: $k=1$ (red curve), $2$ (blue curve), and $3$ (green curve), with the latter value corresponding to the one for which the cosmologically coupled BHs mimic dark energy. \textit{Right panel}: contour plot of the logarithmic amplitude shift $\Delta\log_{10}A$ as a function of $z_i$ and $k$.}
\label{fig:deltalog10a}
\end{figure*}

We now proceed to study the impact of the cosmological coupling on the amplitude of the timing residuals power spectral density signal. To simplify the discussion, we assume that $z_{\min} \to 0$ and $z_{\max} \simeq z_i$. In other words, this means that the coupling is effective only in the redshift range of interest to PTAs (however, we stress that this does not alter our results since, even if $z_i \neq z_{\max}$, the contribution to $\Omega_{\text{gw}}$ from $z_{\max}<z<z_i$ would be common to both the coupled and uncoupled scenarios). Moreover, since as shown by Phinney~\cite{Phinney:2001di} the resulting signal is very weakly dependent on the merger rate $\dot{n}(z)$, for simplicity we assume $\dot{n}(z)=$constant, and assume that the constant is the same in both the cosmologically coupled and standard scenarios. With these approximations in hand, we now quantify the effect of the cosmological coupling on logarithm of the amplitude of $\Phi(f)$. We therefore consider the logarithmic amplitude shift which, inspecting Eq.~(\ref{eq:amodified}), we define as follows:
\begin{eqnarray}
\Delta\log_{10}A(k,z_i)\equiv\frac{5k}{6}\log_{10}(z_i+1)+ \frac{1}{2}\log_{10} \int_{0}^{z_i}\frac{dz}{E(z)(1+z)^{(5k+4)/3}}-\frac{1}{2}\log_{10} \int_{0}^{z_i}\frac{dz}{E(z)(1+z)^{4/3}}\,.
\label{eq:deltalog10a}
\end{eqnarray}
Within the adopted approximations, $\Delta\log_{10}A$ quantifies the logarithmic difference between the amplitude of $\Phi(f)$ in the cosmologically coupled scenario versus the standard case. We work within the assumption of a spatially flat $\Lambda$CDM model, in excellent agreement with current data, so that $E(z)\approx\sqrt{\Omega_m(1+z)^3+(1-\Omega_m)}$ with $\Omega_m \approx 0.3$. We show the resulting logarithmic amplitude shift $\Delta\log_{10}A(k,z_i)$ in Fig.~\ref{fig:deltalog10a}, focusing on the range $k \in [0,3]$: in particular, the left panel shows $\Delta\log_{10}A(k,z_i)$ versus $z_i$ for different values of $k=1$ (red curve), $2$ (blue curve), as well as the maximum allowed value $3$ (green curve), whereas the right panel shows contours of constant $\Delta\log_{10}A$ as a function of both $k$ and $z_i$. Obviously the $k=0$ case, which implies no cosmological growth, corresponds to $\Delta\log_{10}A(k,z_i)=0$.

We see that for $k>0$, the amplitude of the SGWB signal, and with it that of the timing residuals power spectral density signal, both increase. This can be understood in terms of the cosmological mass growth enhancing the energy lost to gravitational radiation [see the dependence on ${\cal M}$ of Eq.~(\ref{eq:dedfr})], which therefore increases the strength of the resulting GW signal. For instance, we see that for $z_i \lesssim 2$ (consistent with the expectations for when DE starts to dominate), the signal can be amplified by up to a factor of $\simeq 1.8$ for $k=1$, $\simeq 3$ for $k=2$, and $\simeq 8$ for the theoretically favored case $k=3$. We also observe that the amplification factor grows more slowly as $z_i$ increases and $k$ decreases.

The prediction that cosmologically coupled (mass-increasing) BHs increase the amplitude of the SGWB signal is particularly intriguing in light of the SGWB signal for which evidence was recently provided by PTA experiments~\cite{NANOGrav:2023gor,EPTA:2023fyk,Reardon:2023gzh,Xu:2023wog}. As already hinted to in the Introduction, the signal is broadly consistent with that predicted from inspiraling supermassive BH binaries, but its amplitude is slightly too large to be explainable by known astrophysical models with reasonable parameters, as recently emphasized by several works~\cite{NANOGrav:2023hfp,Sato-Polito:2023gym,Padmanabhan:2024nvv}: indeed, an explanation of the signal in terms of a population of standard inspiraling supermassive BHs requires either several astrophysical parameters to be at the edges of their expected values, or a few among these parameters to be extremely far from such values. It has been shown that, for standard (non-cosmologically coupled) BHs, there is an upper limit to how much the signal can be amplified by modifications to the merger history, and that the strength of the observed signal appears to require a factor of $\approx 10$ more BHs contributing to the PTA signal compared to standard expectations~\cite{Sato-Polito:2023gym}.

\begin{figure*}[!t]
\centering
\includegraphics[width=0.49\linewidth]{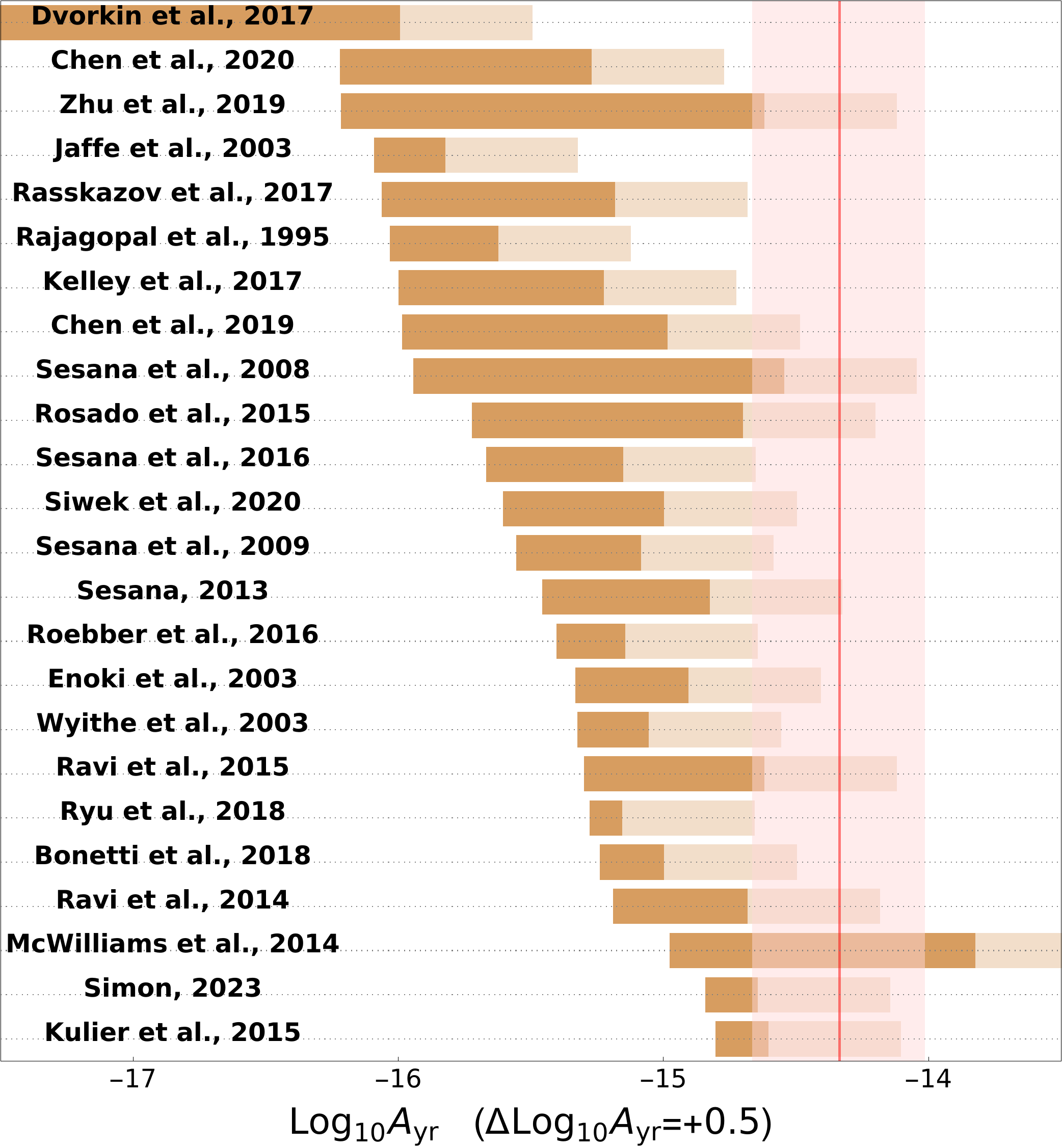}
\includegraphics[width=0.49\linewidth]{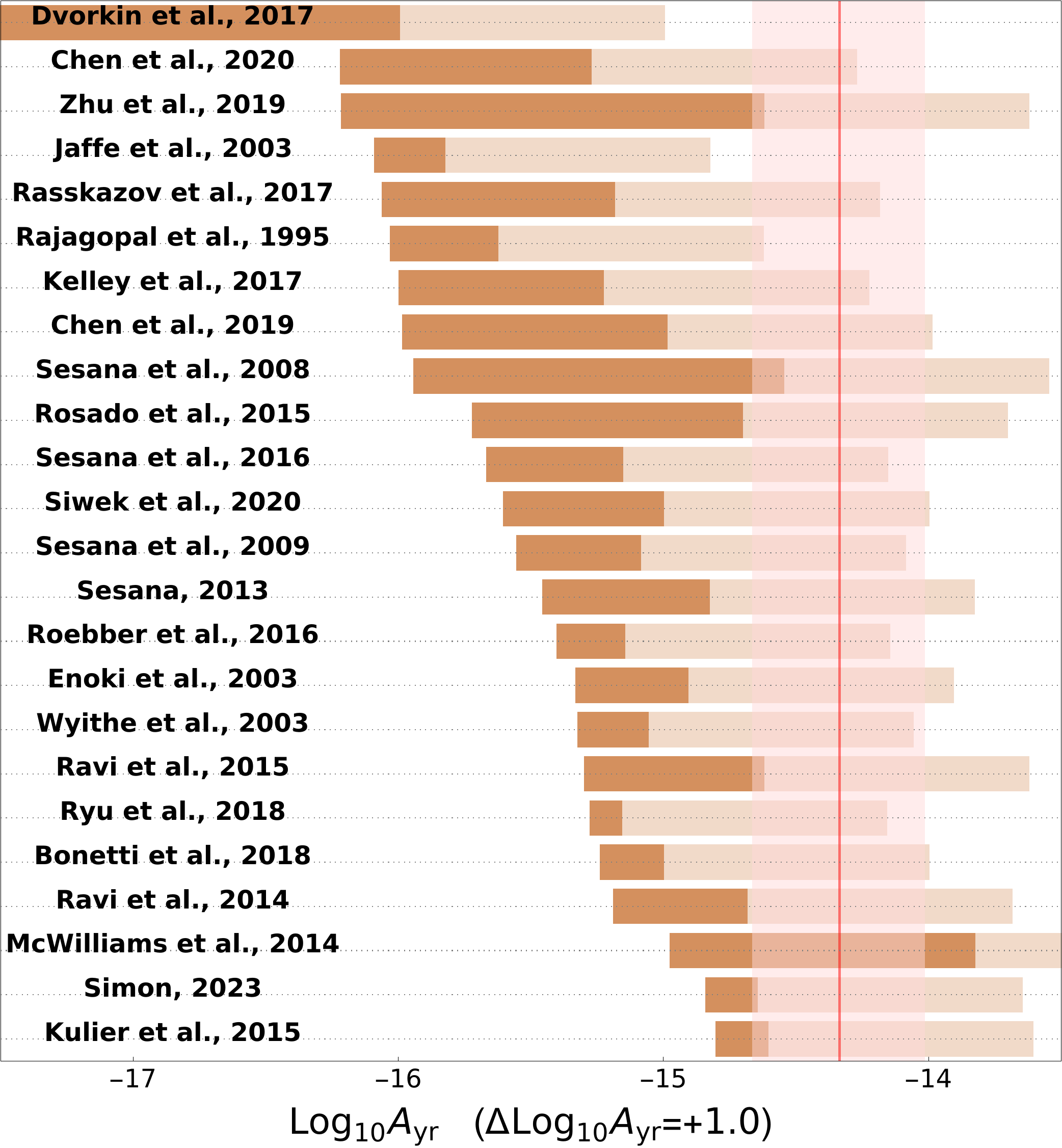}
\caption{Solid horizontal bars: selection of literature predictions for the amplitude of the SGWB from inspiraling supermassive BHs $A_{\text{yr}}$ (amplitude at the reference frequency $1\,{\text{yr}}^{-1}$)~\cite{Rajagopal:1994zj,Jaffe:2002rt,Wyithe:2002ep,Enoki:2004ew,Sesana:2008mz,Sesana:2008xk,Sesana:2012ak,McWilliams:2012an,Ravi:2014aha,Kulier:2013gda,Ravi:2014nua,Rosado:2015epa,Roebber:2015iva,Sesana:2016yky,Rasskazov:2016jjk,Dvorkin:2017vvm,Kelley:2017lek,Ryu:2018yhv,Bonetti:2017lnj,Zhu:2018lif,Chen:2018znx,Chen:2020qlp,Siwek:2020adv,Simon:2023dyi}, with the bars reporting 16$^{\text{th}}$-84$^{\text{th}}$ percentile uncertainty regions. Red vertical shaded region: 68\% credible interval for the SGWB amplitude inferred from a power-law fit to the NANOGrav 15-year data using the HD-DMGP model, with the red vertical line indicating the central value. It is clear that most predictions fall slightly short of the NANOGrav signal. The more transparent parts of the brown horizontal bars instead indicate values of $A_{\text{yr}}$ which can be achieved in the presence of the enhancement due to cosmologically coupled BHs studied in this work, for the representative values $\Delta\log_{10}A_{\text{yr}}=0.5$ (\textit{left panel}) and $\Delta\log_{10}A_{\text{yr}}=1.0$ (\textit{right panel}). The transparent bars represent a constant length augmentation to all the literature predictions for standard supermassive BH models: this treatment is only valid if the model parameters for standard and cosmologically coupled BHs remain unaltered. The Figure should therefore only be interpreted as a qualitative visual guide.}
\label{fig:ayrliteraturepredictionsnanograv}
\end{figure*}

To better appreciate the (slight but nevertheless noticeable) mismatch between theoretical expectations for the signal strength and its observed value, in the bar plot of Fig.~\ref{fig:ayrliteraturepredictionsnanograv} we plot a selection of literature predictions for the amplitude $A_{\text{yr}}$~\cite{Rajagopal:1994zj,Jaffe:2002rt,Wyithe:2002ep,Enoki:2004ew,Sesana:2008mz,Sesana:2008xk,Sesana:2012ak,McWilliams:2012an,Ravi:2014aha,Kulier:2013gda,Ravi:2014nua,Rosado:2015epa,Roebber:2015iva,Sesana:2016yky,Rasskazov:2016jjk,Dvorkin:2017vvm,Kelley:2017lek,Ryu:2018yhv,Bonetti:2017lnj,Zhu:2018lif,Chen:2018znx,Chen:2020qlp,Siwek:2020adv,Simon:2023dyi} (solid horizontal bars), confronted against the amplitude of the NANOGrav 15-year signal~\cite{NANOGrav:2023hfp} (red vertical shaded region). We stress that the horizontal bars are \textit{theoretical predictions} constructed out of model populations for supermassive BH binaries, with specific assumptions on the allowed range of values for the population model parameters, and as such they can be directly compared against the NANOGrav signal. In fact, a similar visual comparison has also been performed by the NANOGrav collaboration (see Fig.~A1 of the NANOGrav supermassive BH binaries paper~\cite{NANOGrav:2023hfp}, with theoretical predictions taken from Tab.~A1 thereof). As already noted by the NANOGrav collaboration, the signal is not dramatically inconsistent with theoretical predictions, although one cannot help but notice that essentially all of these fall short of the observed amplitude (typically by $\lesssim 2\sigma$, albeit in some cases by a larger amount).

For purely representative purposes, in Fig.~2 we show the impact on this discrepancy of an increase in the SGWB amplitude. This could arise within the cosmological coupling scenario described here. To do so, we augment, with increased transparency, the length of the bars corresponding to the theoretical predictions in Fig.~\ref{fig:ayrliteraturepredictionsnanograv} by $\Delta\log_{10}A=0.5$ (left panel) and $\Delta\log_{10}A=1.0$ (right panel), corresponding respectively to half an order of magnitude and an order of magnitude increase in the amplitude of the SGWB signal. We stress that these are chosen purely as benchmark values, and the goal of this exercise in the context of our exploratory work is to visually and qualitatively guide the reader towards the range of values of $\Delta\log_{10}A$ which can reconcile these (otherwise slightly discrepant) literature predictions with the NANOGrav signal. In other words, Fig.~\ref{fig:ayrliteraturepredictionsnanograv} should be seen as the qualitative first version of a statistically more rigorous analysis, which we reserve to future work. Nevertheless, while chosen for representative purposes, these values of $\Delta\log_{10}A$ can be obtained for realistic values of the cosmological coupling parameters, as can be seen in Fig.~\ref{fig:deltalog10a}: for example, $\Delta\log_{10}A=1.0$ can be obtained if $k=3$ and $z_i \gtrsim 2.3$, whereas $\Delta\log_{10}A=0.5$ can be obtained for a very broad range of values of $k \gtrsim 1$ and $z_i \gtrsim 1$. We also stress that our treatment of adding $\Delta\log_{10}A$ to the existing literature predictions is valid only if the assumed population model parameters (BH mass function, merger rate, and so on) are unaltered by the cosmological coupling. Such an assumption, while at first glance reasonable, is actually far from obvious, and requires much more detailed follow-up work going well beyond the scope of our exploratory study.

Returning to Fig.~\ref{fig:ayrliteraturepredictionsnanograv}, we observe at a qualitative level that an increase in the SGWB amplitude at the levels considered therein would be sufficient to bring most predictions from theoretical models in good agreement with the observed strength of the signal, although we reserve a statistically more rigorous analysis to future work. In fact, even for $\Delta\log_{10}A=0.5$, only for three models~\cite{Dvorkin:2017vvm,Jaffe:2002rt,Rajagopal:1994zj} the 16$^{\text{th}}$-84$^{\text{th}}$ percentile uncertainty regions for the theoretical predictions fall significantly outside of the NANOGrav signal, whereas for other three models~\cite{Chen:2020qlp,Rasskazov:2016jjk,Kelley:2017lek} the discrepancy is very mild: while cautioning that an assessment based on Fig.~\ref{fig:ayrliteraturepredictionsnanograv} can only be of qualitative nature, it is clear already at this level that the cosmological coupling mechanism can potentially bring currently discrepant theoretical predictions into better agreement with observations. Fig.~\ref{fig:ayrliteraturepredictionsnanograv} clearly highlights that there is significant scatter across the predictions of astrophysical models for the SGWB signal from inspiraling supermassive BH binaries, and that there is clearly the need for an improved consensus on the topic even in the standard case: nevertheless we can safely draw the conclusion that, should a consensus model confirm the present trend and fall short of the signal observed by NANOGrav, a model featuring cosmologically coupled BHs whose mass increases with the expansion of the Universe can be invoked to address the discrepancy (in addition, of course, to other new physics scenarios, either astrophysical or cosmological in origin).

It is also worth noting that, besides at the level of amplitude, the NANOGrav signal is moderately inconsistent with the theoretical expectations for inspiraling supermassive BH binaries also at the level of spectral index $\gamma$. Indeed, $\gamma=3.2 \pm 0.6$ has been inferred from the observed signal, nearly $2\sigma$ away from the theoretical expectation $\gamma=13/3$~\cite{NANOGrav:2023hvm}. With the caveat that the inference of $\gamma$ is very sensitive to minor details in the data model of a few pulsars, we stress that the theoretical prediction $\gamma=13/3$ is an idealized one, and that environmental effects will generally cause $\gamma$ to deviate from this value (moving towards smaller values). Such effects include but are not limited to interactions with the binary environment~\cite{Begelman:1980vb,Kocsis:2010xa}, the discreteness of the binary population as well as the underlying assembly scenario~\cite{Sesana:2008mz}, orbital eccentricity~\cite{Enoki:2004ew,Enoki:2006kj,Sesana:2013wja,Chen:2016zyo}, dynamical friction, and so on~\cite{Ellis:2023dgf,Ellis:2023oxs}. These effects will of course be relevant in the cosmologically coupled scenario as well.

We close our critical discussion mentioning a few caveats to our analysis, besides all the approximations extensively highlighted earlier. Firstly, recall our assumption that $\dot{n}(z)$ takes the same (constant) value in both the cosmologically coupled scenario and the standard one. This assumption may require revisiting, as it is in principle possible that the increasing cosmologically coupled masses may lead to enhanced clustering of the cosmologically coupled supermassive BHs, potentially increasing the comoving merger rate $\dot{n}$~\cite{Ghodla:2023iaz}. In addition, as stressed above, environmental effects which are expected to alter $\gamma$ from its idealized value of $13/3$ in the standard scenario will certainly be operative in the cosmologically coupled case as well. Although we are far from being able to model and study these effects from first-principles in the case at hand, it is not unreasonable to expect that they might alter the spectral index of the SGWB signal to the same extent in both the standard and cosmologically coupled scenarios (i.e.\ that astrophysical and cosmological effects ``decouple'', though this is not necessarily obvious), and similar considerations hold for environmental effects which alter the amplitude of the SGWB signal -- nevertheless, we reserve a more detailed investigation to follow-up work.

In addition, we have assumed that the expression for $dE_{\text{gw}}/df_r$, itself following from the quadrupole formula, maintains its functional form in terms of $f_r$, with amplitude modified by ${\cal M} \to {\cal M}(z)$. This tacitly assumes that the cosmological coupling and mass growth mechanism do not alter the quadrupole formula. The validity of such an assumption, however, is less obvious than the earlier ones (given that additional terms involving time derivatives of the masses, which are absent in the standard case, may enter, although we can reasonably expect them to be negligible with respect to the term we considered~\cite{Ghodla:2023iaz}). To fully explore the impact of such an assumption requires studying the problem of a varying-mass binary merger, which is already highly non-trivial in the Newtonian limit. In principle, modifications to $dE_{\text{gw}}/df_r$ of this sort could alter the $f^{2/3}$ and $f^{-13/3}$ dependencies of $\Omega_{\text{gw}}(f)$ and $\Phi(f)$ respectively, potentially competing with the aforementioned astrophysical effects. Detailed investigations of this point are ongoing, and we hope to report on the matter in future work. Moreover, while we have assumed the standard $\Lambda$CDM cosmological model with the (effective) DE component in the form of a cosmological constant when specifying $E(z)$ in Eq.~(\ref{eq:deltalog10a}), it has recently been pointed out that the shape of the expansion history in the cosmologically coupled scenario is in fact different, as a result of the effective DE component being ``fed'' by the baryonic one~\cite{Croker:2024jfg}. This leads to an effective time-varying DE model, which has been argued to be compatible with the DESI results~\cite{DESI:2024mwx}. Although the specific cosmological model is not expected to play a major role in determining the predicted signal~\cite{Phinney:2001di}, it may be interesting to make our analysis more realistic by considering either the phenomenological parametrization recently adopted to model the cosmologically coupled scenario~\cite{Croker:2024jfg}. Finally, while here we have focused on positive values of $k$, other works have argued that negative values of $k$ may in fact emerge from different theoretical scenarios~\cite{Cadoni:2023lum}. We have checked that considering values of $-3 \leq k<0$ (with the lower limit set again by causality requirements~\cite{Croker:2021duf}), leading to \textit{decreasing} BH masses, results in a SGWB of lower amplitude, worsening the discrepancy with the PTA signal.

\section*{Conclusions}
\label{sec:conclusions}

The possibility that black holes are cosmologically coupled has recently received renewed attention in both the theoretical and observational communities, and is a relatively robust prediction of scenarios where BH solutions are embedded within cosmological boundary conditions with a time-dependent geometry. The associated surprising consequence of cosmological mass growth, recently argued to be potentially consistent with observations of supermassive BHs within elliptical galaxies, could play a key role in the dark energy phenomenon. Such a cosmological mass growth would also lead to other signatures. In particular, cosmologically coupled supermassive BHs will inspiral and undergo binary mergers throughout their history, leading to a stochastic background of gravitational waves. Motivated by the recent SGWB for which evidence was recently provided by various PTAs and the renewed interest in cosmologically coupled BHs, in this pilot study we have taken a first step towards understanding how the SGWB signal produced by a population of inspiraling cosmologically coupled BHs differs from that predicted within the standard uncoupled scenario.

We have demonstrated that a population of inspiraling supermassive BHs undergoing cosmological mass growth would enhance the amplitude of the resulting SGWB signal relative to the standard case. In a nutshell, such an effect arises from the (cosmological coupling-driven) increase in time of the chirp mass which controls the strength of the emitted gravitational radiation. The amount by which the signal is amplified can be quite large, depending on the coupling strength and redshift at which the coupling becomes effective (which we have quantified by the parameters $k$ and $z_i$), potentially exceeding an order of magnitude (see Fig.~\ref{fig:deltalog10a}). This prediction is of significant interest in light of the PTA signal, whose amplitude exceeds theoretical predictions by a similar margin (see Fig.~\ref{fig:ayrliteraturepredictionsnanograv} for a qualitative visual guide). Although several astrophysical and environmental effects are at play and are, at present, not completely understood, our results open up the intriguing possibility that cosmologically coupled BHs may be at the origin of the signal observed by PTAs.

We stress that our work represents only a first, exploratory step towards understanding the features of the SGWB signal resulting from a population of inspiraling cosmologically coupled BHs, and more work is required to further understand such a signal. There are several other interesting potential follow-up directions one could pursue beyond our pilot study. The most important follow-up direction is to conduct a statistically rigorous analysis to constrain the cosmological coupling parameters $z_i$ and $k$ against PTA data. Such a work entails among other things understanding whether the model parameters for standard and cosmologically coupled BHs can be taken to be identical, and will ultimately lead to a statistically rigorous version of Fig.~\ref{fig:ayrliteraturepredictionsnanograv}, which at present can only be a qualitative visual guide. On the more theoretical side, a form of vacuum energy similar to that driving the present-day accelerated expansion presumably drove inflation: could a similar cosmological coupling mechanism be at play then, leading to a population of cosmologically coupled primordial regular BHs of potential relevance to the dark matter problem? While we plan to return to these and other interesting related questions in follow-up work, including devising novel orthogonal probes of cosmologically coupled BHs we note that the plethora of BH-related observations available to us at the time of writing open up the exciting prospect of testing frameworks which belonged to the realm of theoretical conjectures up to a few years ago~\cite{Barack:2018yly,Odintsov:2022cbm,Vagnozzi:2022moj}. This includes scenarios such as the one studied here, which very naturally merges two among the most important open problems in physics: the nature of dark energy and the singularity problem.

\section*{Acknowledgements}

M.C., M.R., and S.V.\ acknowledge support from the Istituto Nazionale di Fisica Nucleare (INFN) through the Commissione Scientifica Nazionale 4 (CSN4) Iniziativa Specifica ``Quantum Fields in Gravity, Cosmology and Black Holes'' (FLAG). M.C.\ and S.V.\ acknowledge support from the University of Trento and the Provincia Autonoma di Trento (PAT, Autonomous Province of Trento) through the UniTrento Internal Call for Research 2023 grant ``Searching for Dark Energy off the beaten track'' (DARKTRACK, grant agreement no.\ E63C22000500003). This publication is based upon work from the COST Action CA21136 ``Addressing observational tensions in cosmology with systematics and fundamental physics'' (CosmoVerse), supported by COST (European Cooperation in Science and Technology).

\section*{Author contributions statement}

M.R.\ and S.V.\ conceived the project, M.R.\ carried out the analytical calculations, M.C.\ and S.V.\ carried out the numerical calculations, M.C.\ and S.V.\ produced the plots, M.C., M.R., and S.V.\ wrote the initial draft. All authors reviewed and contributed to writing the manuscript. 

\section*{Additional information}

The authors declare no competing interests.

\section*{Data availability}

The datasets used and/or analysed during the current study available from the corresponding author on reasonable request.

\bibliography{sgwb}

\end{document}